# Thermal Boundary Conductance of Two-Dimensional MoS$_2$ Interfaces

Saurabh V. Suryavanshi[1], Alexander J. Gabourie[1], Amir Barati Farimani[2], and Eric Pop[1,3,4,*]

[1]Department of Electrical Engineering, Stanford University, Stanford, CA 94305, USA
[2]Department of Mechanical Engineering, Carnegie Mellon University, PA 15213, USA
[3]Department of Material Science & Engineering, Stanford University, Stanford, CA 94305, USA
[4]Precourt Institute for Energy, Stanford University, Stanford, CA 94305, USA
*Contact: *epop@stanford.edu*

## Abstract

Understanding the thermal properties of two-dimensional (2D) materials and devices is essential for thermal management of 2D applications. Here we perform molecular dynamics simulations to evaluate both the specific heat of MoS$_2$ as well as the thermal boundary conductance (TBC) between one to five layers of MoS$_2$ with amorphous SiO$_2$ and between single-layer MoS$_2$ and crystalline AlN. The results of all calculations are compared to existing experimental data. In general, the TBC of such 2D interfaces is low, below ~20 MWm$^{-2}$K$^{-1}$, due to the weak van der Waals (vdW) coupling and mismatch of phonon density of states (PDOS) between materials. However, the TBC increases with vdW coupling strength, with temperature, and with the number of MoS$_2$ layers (which introduce additional phonon modes). These findings suggest that the TBC of 2D materials is tunable by modulating their interface interaction, the number of layers, and finding a PDOS-matched substrate, with important implications for future energy-efficient 2D electronics, photonics, and thermoelectrics.

I. INTRODUCTION

Two-dimensional (2D) semiconductors such as $MoS_2$ have drawn significant interest for electronic, photonic, and thermoelectric applications.[1,2] In addition, due to their weak van der Waals (vdW) bonds with the environment, they can be directly grown or transferred onto many substrates without the requirement for lattice matching. However, it is this same vdW bond which leads to poor thermal coupling between the 2D material and its environment. This is a particular problem, for example, in three-dimensional (3D) heterogeneous integration scenarios,[3] where high power densities and low thermal conductivity materials create additional thermal dissipation challenges.[1,4] Higher device operating temperatures increase electron-phonon scattering, degrade device performance,[5,6] cause reliability concerns and potential failure.[7]

In this study, we use atomistic molecular dynamics (MD) simulations, compared to existing experimental data,[4,8] to examine in detail the thermal boundary conductance (TBC) between (one to five layers) $MoS_2$ with $SiO_2$ and between single layer $MoS_2$ and AlN substrates. This TBC is particularly important for such 2D materials,[3,4] which have almost no "bulk" and are entirely limited by their interfaces. In particular, recent experiments have found that the interface between $MoS_2$ and $SiO_2$ exhibits a high thermal (Kapitza) resistance, equivalent to that of ~90 nm of $SiO_2$,[4] which is among the highest thermal resistances for solid-solid thermal interfaces.[9,10] However, a detailed theoretical understanding of this thermal interface is presently lacking, an important ingredient necessary for further development and optimization of 2D nanoelectronics.

Established theories of interfacial heat transport, including the acoustic mismatch model (AMM) and the diffusive mismatch model (DMM), can model phonon transport between bulk, 3D materials.[11] In addition to the standard models, there has been extensive work in modeling interfacial heat transport in bulk materials employing MD as well first-principles simulations.[12–15] For 2D materials, one side of the interface is atomically thin and therefore does not necessarily follow the bulk behavior. In addition, most continuum models and first principles calculations for 2D materials do not capture the anharmonicity of phonon coupling across such interfaces.[16–18]

Here we use the LAMMPS package[19] for MD simulations, which accurately captures the phonon scattering physics and anharmonicity of the 2D material and 3D substrate. Heat transport across two electrically insulating materials is dominated by phonons, which can be accurately captured by MD simulations.[20] The calculation of TBC using MD is often done by non-equilibrium MD (NEMD) which requires creating sandwich structures or applying heat sources away from the interface.[21,22] In such structures, hot and cold thermostats are applied on opposing ends and the TBC is calculated based on the temperature drop at the interface. However, such structures cannot always replicate the experimental devices without a superstrate,[4] which could affect the interfacial heat flow.[23] In this work, we employ the approach to equilibrium method (AEMD) which heats up the 2D material and then cools it down through the substrate, consistent with certain experiments.[4]

## II. SIMULATION DETAILS

The simulation box contains one to five layer(s) of $MoS_2$ on an insulating substrate such as amorphous $SiO_2$ or crystalline AlN, as illustrated in Fig. 1(a). We chose $MoS_2$ as a representative 2D material[24] because experimental data of its thermal interface have recently become available.[4,8,22] However, we directly compare our simulations only to Refs. [4,8] because Ref. [22] had a Ti interface, which is known to react with $MoS_2$[25] and no longer forms a vdW bond.

To model the interatomic interactions, we use the Tersoff potential parameterized by Munetoh *et al.* for $SiO_2$[26] and a modified Stillinger-Weber potential (SW) for AlN by Vashishta *et al.*[27] For $MoS_2$, we corrected the SW potential originally parameterized by Jiang *et al.*[28] (See Supplement section 1 for the corrections implemented.) These potentials have been previously used to predict the thermal properties of their respective materials.[24,27–29] We treat the non-bonding interactions between the $MoS_2$ layer and the substrate using the 12-6 Lennard-Jones (LJ) potential given by $V(r) = 4\chi\varepsilon[(\sigma/r)^{12} - (\sigma/r)^6]$ where $\varepsilon$ is the energy parameter, $\sigma$ is the distance parameter, $r$ is the interatomic distance, and $\chi$ is the scaling factor for the interaction energy.[21] Because there is some uncertainty regarding the interaction potential between $MoS_2$ and the substrate, we later scale $\chi$ (from 1 to 4) to study the dependence of TBC on the vdW interaction strength. The parameters used in the LJ potential are obtained by applying mixing rules to the universal force field (UFF)[30] parameters (see Supplement section 2 for additional details).

The $MoS_2$ layer(s) are placed either on an amorphous $SiO_2$ substrate as shown in Fig. 1(a), or on a crystalline AlN block, as shown in supplementary Fig. S1. The Supplement (section 3) describes the equilibration procedure of such systems, necessary to obtain realistic results consistent with the experimental data. To simulate heat transfer across the $MoS_2$-$SiO_2$ interface, we create an initial temperature difference $\Delta T_0(t=0) = T_{MoS2} - T_{SiO2} \approx 200$ K between the two materials by applying two separate Nosé-Hoover thermostats.[29] The initial temperature conditions are similar for AlN substrate. Such a temperature difference is consistent with experimental measurements of MoS2 transistors, when biased in DC operating conditions.[4] We simulate both MoS2 and the substrate at a constant volume and a constant temperature (NVT) with a time step of 0.1 fs for 1 ns to allow them to reach equilibrium. We then switched the MoS2 and top four-fifths of the substrate to constant energy and constant volume ensemble (NVE) with a time step of 0.05 fs. The time step is decreased to capture high frequency phonons. The bottom one-fifth of the substrate is maintained at a constant temperature (NVT) $[T_{SiO2}(t=0)]$ as a boundary condition.[31,32,33] The application of an NVT ensemble to the bottom one-fifth of the substrate rescales the atomic velocities and reduces the phonon reflection from the back surface. The MoS2 layer cools by dissipating heat to the substrate through the interface, similar to heat dissipation in realistic devices.[4]

We visualize the spatial and temporal variation of temperature of the top ~25 Å of the MoS2-SiO2 structure in Fig. 1(b). Each box in this figure represents the temperature of a ~2 Å section averaged over 10 ps. The local temperature is calculated from the local atomic masses and velocities. In Fig. 1(c), we plot the temperature of the MoS2 layer (all Mo and S atoms) and SiO2 block (all Si and O atoms from the upper four-fifths of the substrate) as a function of time. The MoS2 temperature is drastically reduced and the SiO2 temperature only increases slightly because it has a much larger heat capacity. Because MD simulations are deterministic, we perform eight independent runs for each setup (temperature, χ, substrate), with different initial atomic velocity distributions, to calculate standard deviations.

To extract the TBC, we fit the time decay of the temperature difference in Fig. 1(c) to the thermal resistive-capacitive (RC) network shown in Fig. 1(d) as

$$\Delta T = \Delta T_0 e^{-\frac{t}{\tau}} = \Delta T_0 e^{-\left(\frac{1}{m_{MoS2}C_{MoS2}} + \frac{1}{m_{SiO2}C_{SiO2}}\right)GAt}. \quad (1)$$

Here, $\tau$ is the characteristic decay time, $C_{MoS2}$ and $C_{SiO2}$ are the specific heat per unit mass of MoS2 and SiO2, $m_{MoS2}$ and $m_{SiO2}$ are the total masses of the MoS2 layer and the top four-fifths of the SiO2 block respectively, $A$ is the area of the interface between MoS2 and the substrate, and $G$ is the TBC. Since $m_{SiO2} \gg m_{MoS2}$, the temperature decay is primarily dominated by $C_{MoS2}$ and $G$. We include quantum corrections[34] to calculate the heat capacity of a single layer (1L) of MoS2 as

$$C_{MoS2} = 3Nk_B \frac{\int_0^{\omega_{max}} \frac{u^2 e^u}{(e^u - 1)^2} D(\omega) d\omega}{\int_0^{\omega_{max}} D(\omega) d\omega}, \quad u = \frac{\hbar \omega}{k_B T}. \quad (2)$$

Here, $N$ is the total number of atoms, $k_B$ is the Boltzmann constant, $\omega$ is the phonon frequency, $\omega_{max}$ ($\approx$ 15 THz from our MD simulations for monolayer MoS$_2$) is the maximum phonon frequency, and $D(\omega)$ is the phonon density of states (PDOS), calculated as the Fourier transform of the atomic velocity autocorrelation function

$$D(\omega) = \int_0^\infty \langle \vec{v}(t) \vec{v}(0) \rangle e^{i\omega t} dt \; D(\omega) = \int_0^\infty \langle \vec{v}(t) \vec{v}(0) \rangle e^{i\omega t} dt, \quad (3)$$

where the velocities $v(t)$ are obtained by monitoring the atomic motions directly within the MD simulations, with temperature $T$ at equilibrium (NVE).

Equation (2) allows us to calculate the temperature dependence of the heat capacity from the PDOS of monolayer MoS$_2$, shown in Fig. 2(a). At room temperature we obtain $C_{MoS2} \approx$ 61 Jmol$^{-1}$K$^{-1}$, which compares well to the experimental value of ~63.8 Jmol$^{-1}$K$^{-1}$ for bulk MoS$_2$.[35] Within MD simulations, the heat capacity remains independent of temperature.[36] In our calculations, we use $C_{MoS2}$ = 75 Jmol$^{-1}$K$^{-1}$, which is also high temperature limit ($T \gg$ the Debye temperature of MoS$_2$, $\theta_D$) of Eq. (2). MD simulations follow Maxwell-Boltzmann statistics instead of Bose-Einstein statistic for phonons, limiting their application to temperatures higher than $\theta_D/3$.[9] For monolayer MoS$_2$ $\theta_D \sim$ 262 K[37] (for bulk MoS$_2$ $\theta_D \sim$ 300 K),[35] thus we avoid temperatures below 150 K in our simulations, where MD results are expected to be less accurate.

## III. RESULTS AND DISCUSSIONS

### A. TBC Dependence on Interaction Strength ($\chi$)

Figure 2(b) displays the calculated TBC as a function of substrate temperature and of the MoS$_2$-SiO$_2$ vdW interaction strength, $\chi$. At 300 K, we calculate $\tau \sim$ 130 ps and $G$ = 11.5 ± 0.8 MWm$^{-2}$K$^{-1}$ for monolayer MoS$_2$ on SiO$_2$, with $\chi$ = 1. However, for $\chi$ = 2 we get a higher value of $G$ = 25.6 ± 3.3 MWm$^{-2}$K$^{-1}$ ($\tau \sim$ 65 p). The existing experimental data[4,8] is found to be between simulations values of $\chi$ =1 and $\chi$ = 2 (for higher temperature the data is closer to $\chi$ = 2 results), which suggests that the interaction between $MoS_2$ and $SiO_2$ is likely to be stronger than that estimated by the mixing rules and the UFF model.[30] This is likely because the mixing rules do not accurately consider polarization when two different atoms are brought in contact with each other. Because we model vdW interactions using the 12-6 LJ potential, increasing the interaction strength increases the depth of the potential energy well in the LJ energy landscape. Changing $\chi$, however, does not change the equilibrium distance between the $MoS_2$ layer and the substrate, which was verified by our MD simulations.

A TBC of 14 to 25 $MWm^{-2}K^{-1}$ across the monolayer $MoS_2$-$SiO_2$ interface near room temperature (encompassing the range of experimental data[4,8] and our simulations with $\chi = 2$) is equivalent to the thermal resistance of ~55 to 100 nm thick amorphous $SiO_2$ (the so-called Kapitza length), assuming $SiO_2$ thermal conductivity of 1.4 $Wm^{-1}K^{-1}$. This is a substantial thermal resistance, especially in the context of nanoscale devices. In comparison, the low end of known solid-solid interface TBC is ~8 $MWm^{-2}K^{-1}$ for Bi-diamond, only a factor of two to three lower, while the upper end of known TBC is ~14 $GWm^{-2}K^{-1}$ for Pd-Ir interfaces where electronic heat conduction is significant.[9,10] The TBC of graphene with $SiO_2$ has been measured in the range of 25 to 65 $MWm^{-2}K^{-1}$, weakly proportional to the number of layers.[38,39]

From Fig. 2(b), we also see that the TBC increases almost proportionally with $\chi$, which is not unexpected. Higher interaction strength increases the harmonic coupling between $MoS_2$ and substrate, improving the phonon transmission for heat transfer.[29] By further increasing the interaction strength, the TBC increases with improvement in transmission but eventually becomes mode-limited and begins to saturate at $\chi > 15$ (Fig. S2 in Supplement). This interaction strength dependence highlights an important aspect of modulating TBC. First, any interfacial residue (e.g. polymer resist from fabrication) is likely to reduce the interaction between 2D material and substrate. Therefore, to improve TBC for 2D materials it is important to make cleaner interfaces. Second, it is also possible to increase the TBC by increasing the interaction between 2D material and substrate, for example by functionalizing either of them.[40]

We comment on the apparently higher TBC measurements reported in Ref. [22], up to ~33.5 $MWm^{-2}K^{-1}$ for $MoS_2$ on $SiO_2$. These measurements were performed with a Au/Ti metal heater on top of the $MoS_2$, as a type of "indirect heating," i.e. the heat passing through the $MoS_2$ but not being generated inside it. In addition, Ti is known to react with $MoS_2$,[25] no longer forming a vdW bond. In contrast, Refs. [4,8] and the simulations in this work performed "direct heating," where heat is generated inside the MoS$_2$ and passes across the vdW interface to the SiO$_2$, i.e. the scenario of MoS$_2$ electronic devices during operation. Nevertheless, the *indirect* heating experiments also highlight one of the core messages of our simulations, which is that strengthening the MoS$_2$ bonding with its environment should enable higher TBC. In addition, *direct* heating experiments reveal a lower TBC due to an "internal thermal resistance" between optical phonons, which are partly heated, and the acoustic phonons, which are primarily responsible for carrying heat across the interface.[41,42]

### B. TBC Dependence on Temperature

Figure 2(b) also displays the calculated temperature dependence of the TBC, for varying substrate temperatures. For $T \ll \theta_D$, the TBC follows $\sim T^3$ dependence and for $T \gg \theta_D$ the TBC remains constant.[9,11] For intermediate temperatures, we expect the TBC to follow $\sim T^n$ behavior, where $0 < n < 3$. As expected, we extract $n = 0.26$ to $0.30$ for monolayer MoS$_2$ on SiO$_2$, while the experimentally observed value of $n = 0.65$ in a similar temperature range.[8] Interestingly, we see that the larger vdW interaction strength yield smaller $n$ values. This suggests that interfacial coupling, in addition to the material Debye temperature, plays a role in the temperature dependence of the TBC.

### C. TBC Dependence on Number of 2D Layers

To study the dependence of TBC on the number of 2D layers, we repeat the same methodology described earlier, but replacing the monolayer MoS$_2$ with Bernal-stacked (ABA) layers, in the 2H phase. We fix the initial interlayer distance (distance between Mo atoms in adjacent layers) to 6.15 Å based on neutron scattering studies of bulk MoS$_2$.[43] The vdW interactions between the layers are modeled using the 12-6 LJ potential (see Supplement section 2 for details). After the equilibration step, the MoS$_2$ layers were separated by $\sim 6.82$ Å in the bilayer structure and $\sim 6.18$ Å in the tri-layer structure. As we increase the layer number further, the interlayer distance approaches $\sim 6.15$ Å, similar to the experimentally measured distance in bulk MoS$_2$.[43]

To calculate the *effective* multi-layer TBC with SiO$_2$, we average the temperature ($T_{\text{MoS2}}$) over all MoS$_2$ layers and consider the total mass of the MoS$_2$ layers in Eq. (1). This is necessary because the MoS$_2$ layers start at the same temperature in our simulations, but then equilibrate differently as heat from the top layer must pass through the bottom layers, etc. The *intrinsic* TBC is the inverse of the thermal resistance between the bottom MoS$_2$ layer and the top of the SiO$_2$. However, for a device, we are interested in the effective cooling of the (multi-layer) MoS$_2$ which is accounted by an *effective* TBC. (Note that for 1L MoS$_2$, the effective TBC will be same as the intrinsic TBC.) In Fig. 3(a), we see that this effective TBC increases slightly from 1L to 3L MoS$_2$ and then saturates to $G = 13.0 \pm 0.8$ MWm$^{-2}$K$^{-1}$ (with $\chi = 1$). The larger TBC of the multi-layer MoS$_2$ is due to contribution from additional flexural phonon branches, consistent with previous observations for WSe$_2$ and graphene.[17,23] The effective TBC eventually saturates due to increase in the inter-layer resistance for multi-layer MoS$_2$. We approximate the contribution from the intrinsic TBC in the Supplement Section 5 and notice that this increases with the number of layers, as expected. The temperature relaxation times in Fig. 3(b) increase with the number of MoS$_2$ layers, i.e. linearly with the $m_{MoS2}C_{MoS2}$. In other words, to cool down a five-layer MoS$_2$ device that is only limited by its TBC, would require at least $3\tau$ or ~1.1 ns.

### D. TBC Dependence on Substrate

For transistor applications of 2D materials, the thermal resistance is often dominated by the interfacial resistance (1/TBC) and the thermal resistance of the underlying substrate.[5] Therefore, it might seem beneficial to use a high thermal conductivity (TC) substrate such as AlN instead of amorphous SiO$_2$. However, different substrates have different vibrational modes and PDOS, potentially leading to different phonon coupling and TBC across their interface with MoS$_2$. As an example, the PDOS for SiO$_2$, AlN, and MoS$_2$ are quite different from each other and shown in Fig. 4(a). Knowing that the TBC of 2D materials is dominated by low-frequency ZA (i.e. flexural acoustic) modes,[44] the substrate must have sufficient low-frequency states to enable coupling of these low-frequency 2D phonons. In other words, modeling the low-frequency PDOS is expected to be sufficient to capture the trend of the TBC between the various materials. The application of an NVT ensemble to the bottom of the substrate rescales atomic velocities which reduces spurious phonon reflections and avoids finite size effects in MD simulations.

In this context, Fig. 4(a) reveals that ZA modes of monolayer MoS$_2$ range from 0 to ~5 THz,[45] and that the PDOS for SiO$_2$ is larger than for AlN at low frequencies, consistent with the size of Si and O atoms being larger than Al and N atoms. Given the larger overlap in the low-frequency PDOS for SiO$_2$, our MD simulations find that the TBC with the SiO$_2$ substrate is up to a factor of two larger than the TBC with the AlN substrate [Fig. 4(b)].[18]

These results indicate that high TC alone is insufficient to improve heat sinking from 2D material devices, and that it could be important to modify the substrate surface to match the PDOS with the 2D material. For example, we note that among experiments, AlN tends to form a native oxide layer;[46] thus, the presence of heavier O atoms at the surface could increase the low-frequency PDOS, leading to an experimentally measured TBC that is similar to that with $SiO_2$.[8]

### E. TBC and transistor self-heating

Before concluding, we comment on the role of TBC in the context of $MoS_2$ transistors, where recent measurements[4,6] have shown that electrical performance is strongly limited by self-heating during operation (e.g. in the high-field, high-current regime), as they represent an extreme case of semiconductor-on-insulator (SOI) technology.[7,47] To avoid detrimental self-heating, such transistors must either be placed on "perfect heat-sink" substrates (e.g. with well-matched PDOS, high TBC, and high TC), or operated on time scales comparable to the thermal transient of the 2D material, $\tau$, calculated above. In other words, transistor heating can be minimized if they are switched "on" for time scales $t_{on} < \tau$, or if they are switched off for time scales $t_{off} \gg \tau$ (e.g. 1 ns). The MD simulations above put a fundamental lower limit on these thermal time constants, limited only by the heat capacity of $MoS_2$ and the TBC with the substrate. In practice, the thermal time constants of such devices are higher due to the presence of capping oxides and metal contacts, which increase the effective heat capacity.[48]

### IV. CONCLUSIONS

In summary, we have described detailed MD simulations examining the thermal boundary conductance of a 2D material (one to five layers of $MoS_2$) with two technologically relevant insulating substrates, $SiO_2$ and AlN. These TBC values are near the lower limit of TBCs known for solid-solid interfaces,[9,10] due to weak van der Waals bonding and PDOS mismatch between the 2D material and 3D substrate. Nevertheless, the TBC could be increased by maintaining clean interfaces, by strengthening the bond with the substrate, by improving the PDOS matching, or by increasing the number of 2D layers. The latter is also more amenable to modification by stronger chemical bonds (e.g. with Ti), where the topmost layer may be "damaged" by the chemical reaction, but with an overall improved TBC. These results could lead to engineered TBC of 2D materials, e.g. lowering it for thermoelectric devices, and increasing it to reduce self-heating in 2D-based electronics.

## SUPPLEMENTARY MATERIAL

See supplementary material for modification of SW potential for MoS$_2$, calculation of UFF parameters for LJ potential, and details about structural equilibration process. In the supplement, we also include additional discussion regarding the role of the interaction strength ($\chi$) on TBC and provide estimate for intrinsic TBC between MoS$_2$ and SiO$_2$.

## ACKNOWLEDGMENTS

We would like to thank Dr. Eilam Yalon and Dr. Aditya Sood for the constructive discussion and insights into experimental work. This work has been supported by the NCN-NEEDS program, which is funded by the National Science Foundation (NSF) contract 1227020-EEC and by the Semiconductor Research Corporation (SRC). The study was also partly supported by the NSF EFRI 2-DARE grant 1542883, by the Air Force grant FA9550-14-1-0251, and by ASCENT, one of six centers in JUMP, a SRC program sponsored by DARPA. AJG acknowledges partial support from an NDSEG fellowship. We also thank Stanford University and the Stanford Research Computing Center (Sherlock cluster) for providing computational resources and support.

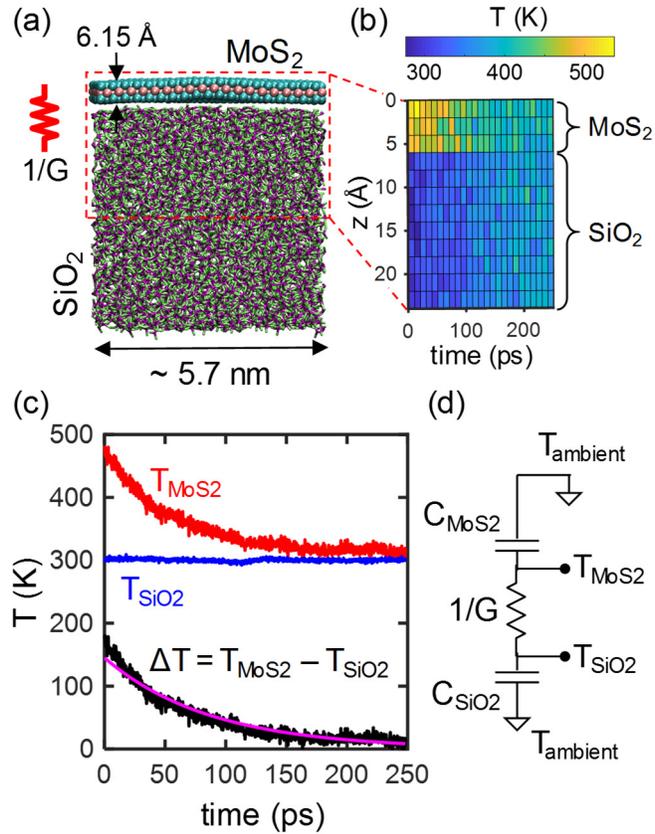

**Figure 1:** Approach to equilibrium molecular dynamics (AEMD) simulations of the $MoS_2$-$SiO_2$ interface. (a) Single layer $MoS_2$ on a $(5.7~nm)^3$ block of $SiO_2$. In-plane periodic boundary conditions are assumed to simulate a continuous sheet of $MoS_2$. $G$ is the thermal boundary conductance between monolayer $MoS_2$ and $SiO_2$. The system is divided into horizontal "slices" (sections) of ~2 Å thickness. The temperature of each slice is averaged over 10 ps, which is represented by a single box. (b) A typical temperature evolution map for 1L $MoS_2$ on the $SiO_2$ system. (c) A typical simulation of temperature transients. The simulation monitors the temperatures of $MoS_2$ (red) and $SiO_2$ (blue) and extracts the TBC from fitting the $\Delta T$ (black) with the thermal RC circuit. (d) Thermal RC circuit of the structure shown in (a) to analyze AEMD simulations.

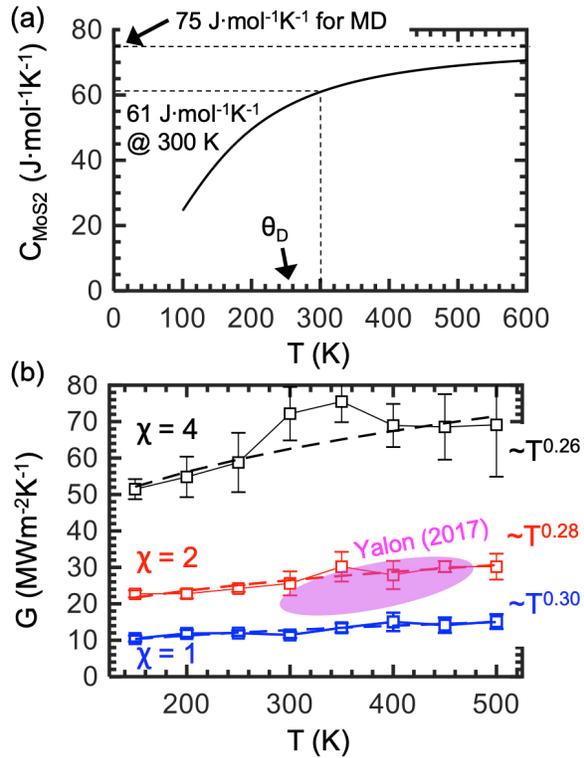

**Figure 2:** (a) The heat capacity ($C_{MoS2}$) calculated for the monolayer (1L) MoS$_2$ from Eq. (2). The Debye temperature ($\theta_D \sim 262$ K) of monolayer MoS$_2$ is also shown. A constant heat capacity of ~75 Jmol$^{-1}$K$^{-1}$ is used for TBC calculations. (b) TBC values from MD simulations, showing the dependence of TBC on the interaction strength $\chi$ and the substrate temperature. The error bars show confidence intervals of one standard deviation. At room temperature and $\chi = 2$, we obtain $G = 25.6 \pm 3.3$ MWm$^{-2}$K$^{-1}$ which is near the experimentally measured ~15 MWm$^{-2}$K$^{-1}$, within the experimental and computational error bars.[8]

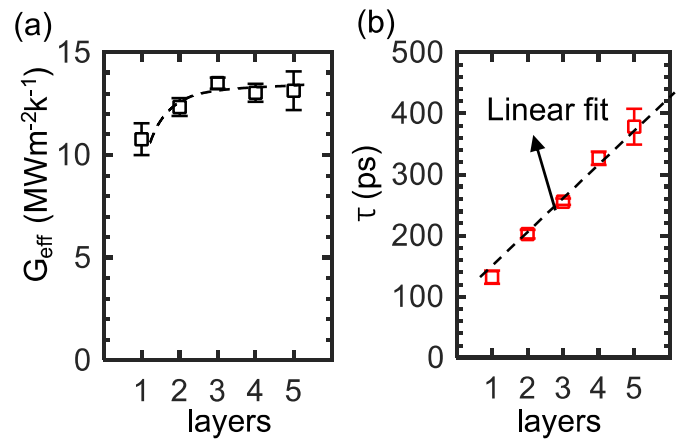

**Figure 3:** (a) Calculated effective TBC as a function of the number of MoS$_2$ layers on SiO$_2$, here with $\chi = 1$. *G* shows a slight increase but saturates to a constant value. (b) The decay time ($\tau$) as a function of the number of layers. $\tau$ shows a linear relationship to the number of layers and the heat capacity of the MoS$_2$ layer(s).

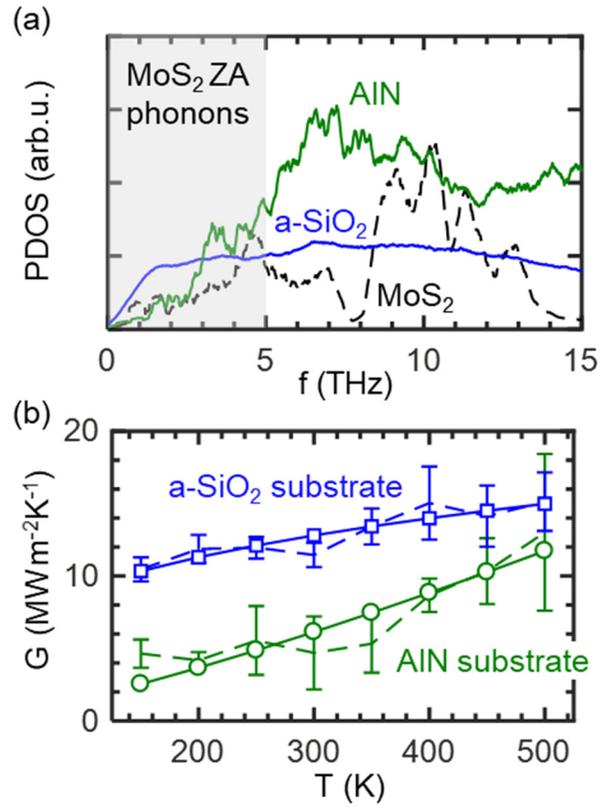

**Figure 4:** (a) Phonon density of states (PDOS) for amorphous $SiO_2$, crystalline AlN, and 1L $MoS_2$. The grey region highlights the phonon frequencies occupied by ZA mode phonons in monolayer $MoS_2$ (b) Calculated temperature dependence of TBC for monolayer $MoS_2$ on amorphous $SiO_2$ and crystalline AlN (with $\chi = 1$).

# Supplementary Information

## Thermal Boundary Conductance of Two-Dimensional MoS$_2$ Interfaces

Saurabh V. Suryavanshi[1], Alexander J. Gabourie[1], Amir Barati Farimani[2], and Eric Pop[1,3,4,*]

[1]Department of Electrical Engineering, Stanford University, Stanford, CA 94305, USA
[2]Department of Mechanical Engineering, Carnegie Mellon University, PA 15213, USA
[3]Department of Material Science & Engineering, Stanford University, Stanford, CA 94305, USA
[4]Precourt Institute for Energy, Stanford University, Stanford, CA 94305, USA
*Contact: *epop@stanford.edu*

### 1. Modified Stillinger-Weber (SW) MoS$_2$ potential

The molecular dynamics (MD) simulations are governed by interatomic potentials. For MoS$_2$, we need to consider three-body interactions in addition to two-body interactions. As a result, the Stillinger-Weber (SW) potential which accounts for both types of interactions has been widely used in the literature to simulate the thermal properties of MoS$_2$.[1] The SW potential, parameterized by Jiang *et al.*,[1] enforces an additional cutoff for S-Mo-S triplets (where Mo is the vertex of the angle formed by the three atoms) primarily to save computation time. With this, the three-body interaction is considered only when the S-S distance ($d$) is greater than the hard-coded cutoff ($d_0$ = 3.78 Å). In order to use this cutoff, the original *pair_sw.cpp* from the LAMMPS source code[2] needs to be replaced by the source code provided by Jiang *et al.* within the supplement of Ref. [1].

In the source code from Ref. [1] the *square* of the distance ($d^2$) between two S-S atoms was incorrectly compared to the additional cutoff ($d_0$ = 3.78 Å) and not the square ($d_0^2$). This caused all three-body interactions to be ignored by default during MD simulations. Simply squaring the additional cutoff ($d_0^2$) resolved this problem and restored the correct three-body interactions. In our earlier preliminary study,[3] we had used the uncorrected potential. By considering this correction in the present study, our computed TBC for single layer MoS$_2$ on SiO$_2$ (at $\chi = 1$) is 11.5 ± 0.8 MWm$^{-2}$K$^{-1}$ instead of 15.5 ± 1.5 MWm$^{-2}$K$^{-1}$, as obtained previously in Ref. [2].

Our corrections to line 88 of *pair_sw*.cpp from Ref. [1] are shown below:

```
replace:      double bondss = 3.78;
with:         double bondss = 14.2884;
```

### 2. Universal force field (UFF) parameters for Lenard-Jones (LJ) potential

We derive the LJ parameters for interfacial interaction from the universal force field (UFF)[4] using the Lorentz-Berthelot mixing rules. The UFF provides the LJ energy ($\varepsilon$) and distance ($\sigma$) parameter for interaction between a pair of *identical* atoms. Following the Lorentz-Berthelot mixing rules, the energy parameter for a pair of dissimilar, non-bonded atoms is obtained by taking a geometric mean of each participating atom's energy parameter. The distance parameter is obtained by taking an arithmetic mean of distance parameters of the participation atoms.[5] In other words, for finding the energy ($\varepsilon_{A-B}$) and distance ($\sigma_{A-B}$) parameter for a pair of atoms A and B, we use the equations:

$$\varepsilon_{A-B} = \sqrt{\varepsilon_{A-A} \cdot \varepsilon_{B-B}} \quad \text{(S1)}$$

$$\sigma_{A-B} = \frac{\sigma_{A-A} + \sigma_{B-B}}{2} . \quad \text{(S2)}$$

The parameters obtained for various pairs are listed in the following table.

| Atomic pair (A and B) | $\varepsilon_{A-B}$ (meV) | $\sigma_{A-B}$ (Å) |
|---|---|---|
| Mo-Si | 6.52 | 3.274 |
| Mo-O | 2.51 | 2.920 |
| S-Si | 14.39 | 3.721 |
| S-O | 5.58 | 3.385 |
| Mo-Al | 7.29 | 3.362 |
| Mo-N | 2.69 | 2.991 |
| S-Al | 16.13 | 3.803 |
| S-N | 5.96 | 3.428 |
| Mo-Mo | 2.40 | 2.720 |
| S-S | 11.88 | 5.396 |
| Mo-S | 5.34 | 3.544 |

**Table S1:** The values for energy and distance parameters required for LJ potentials.

## 3. Structure equilibration process and calculation of TBC

To stabilize the substrates, we first perform a separate equilibration simulation for 200 ps with a time step of 0.01 fs under a constant pressure of 1 bar at 300 K (NPT). A 20 nm vacuum is employed above the substrate to allow its surface to relax and reconfigure. The other two directions employ periodic boundary conditions (PBCs). We then place the MoS$_2$ layer(s) at ~3 Å from the surface and allow the entire structure to relax. The MoS$_2$ has similar periodic boundaries as the underlying substrate. We then minimize the energy by iteratively adjusting the atomic coordinates until the change in energy, for each atom, is less than $10^{-6}$ eV and the force in every direction on every atom is less than $10^{-2}$ eV/Å. We further equilibrate the system for 200 ps with a time step of 0.01 fs under a constant pressure of 1 bar at 300 K (NPT). A stable system for MoS$_2$ on SiO$_2$ is shown in Fig. 1(a) of the main manuscript. Figure S1 below shows a stable system of MoS$_2$ on AlN.

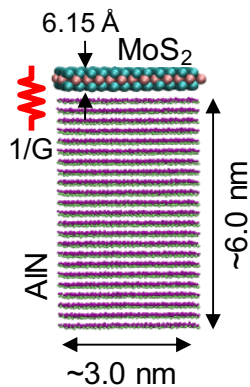

**Figure S1:** Single layer $MoS_2$ on a block of AlN. In-plane periodic boundary conditions are assumed to simulate a continuous sheet of $MoS_2$. $G$ is the thermal boundary conductance between monolayer $MoS_2$ and AlN.

## 4. Role of van der Waals (vdW) interaction strength ($\chi$)

We change the interaction strength ($\chi$) to understand the role of interfacial interaction in heat transfer across the 2D vdW interface. Figure S2 below shows the calculated thermal boundary conductance (TBC) as a function of $\chi$.

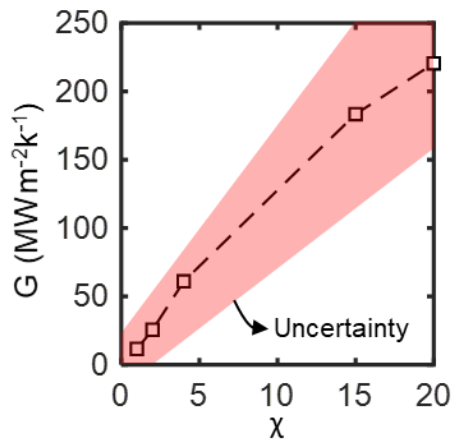

**Figure S2:** Dependence of TBC on the vdW interaction strength ($\chi$) between monolayer $MoS_2$ and $SiO_2$. The dashed black line shows the trend to guide the eyes. The uncertainty region displayed in pink is the standard deviation obtained by performing eight different simulations with different starting atomic velocities.

## 5. Estimation of intrinsic TBC with the number of layers

For multi-layer $MoS_2$ devices, we have focused on the *effective* TBC between the $MoS_2$ and the substrate in the main body of the manuscript. This assumes an average temperature of the $MoS_2$ stack, thus including part of the thermal resistance to heat flow through the (multiple) $MoS_2$ layers.

We have found this saturates to a constant value for stacks thicker than 3 layers (Fig. 3 from main manuscript).

Here we examine the *intrinsic* TBC ($G_i$), which is the inverse of the thermal resistance between the bottom MoS$_2$ layer and the substrate, for a multi-layer MoS$_2$ stack. First, we note that the TBC between two layers of ABAB stacked MoS$_2$ is $G_{\text{MoS2-MoS2}} \approx 44.5$ MWm$^{-2}$K$^{-1}$.[6] For $N$ layers of MoS$_2$, we estimate the intrinsic TBC from the effective TBC ($G$) as:

$$\frac{1}{G_i} = \frac{1}{G} - \frac{N-1}{2} \cdot \frac{1}{G_{MoS2-MoS2}} \quad (S3)$$

Using Eq. S3, we plot $G_i$ as a function of the number of layers in Fig. S3, noting a proportional increase of the intrinsic TBC with number of layers, in this range. Although our calculations are difficult to carry out for thicker MoS$_2$ films ($N > 5$), we expect that $G_i$ will approach a constant, that of the bulk MoS$_2$ TBC with SiO$_2$ in the limit of the film thickness being several times the phonon mean free path for heat flow across MoS$_2$, i.e. in samples thicker than ~140 nm.[7] We also note that this approximation will not change the extracted decay time as shown in Fig. 3(b) in the main manuscript, which is for the entire multi-layer stack of MoS$_2$.

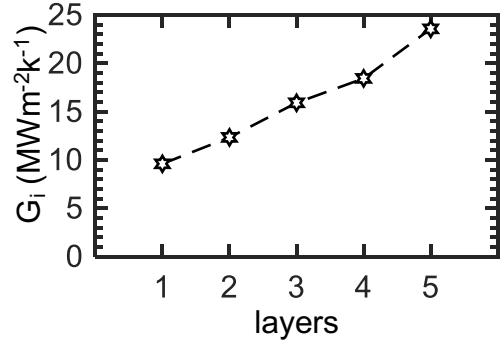

**Figure S3:** Estimated intrinsic TBC of multi-layer MoS$_2$, between the bottom layer and the amorphous SiO$_2$ substrate, with $\chi = 1$.

**Supplementary References:**